# Contribution Rate Imputation Theory: A Conceptual Model

**Vincil Bishop III**
Department of Systems Engineering
Colorado State University
Vincil.Bishop@colostate.edu

**Steven Simske**
Department of Systems Engineering
Colorado State University
Steve.Simske@colostate.edu

## Abstract

The "Theory of Contribution Rate Imputation" estimates developer effort by analyzing historical commit data and typical development rates. Building on the Time-Delta Method, this approach calculates unobserved work periods using metrics like cyclomatic complexity and Levenshtein distance. The Contribution Rate Imputation Method (CRIM) improves upon traditional productivity metrics, offering a more accurate estimation of person-hours spent on software contributions. This method provides valuable insights for project management and resource allocation, helping organizations better understand and optimize developer productivity.

## 1 Introduction

In modern software development, accurately assessing the effort spent on individual contributions is crucial for effective project management, resource allocation, and developer productivity analysis. Despite advancements in development practices, quantifying the time dedicated to specific code contributions remains a challenge due to the distributed, asynchronous, and often collaborative nature of software development. Traditional metrics such as lines of code and commit counts fail to capture the true effort involved in producing meaningful contributions, often overlooking the complexity of the work and the actual time spent.

This paper introduces the Contribution Rate Imputation Method (CRIM) as a novel approach to estimating developer effort through the analysis of historical development data and contribution patterns. CRIM is designed to bridge the gap between traditional contribution metrics and real-world developer behavior, offering a method to impute the time spent on contributions even when direct observations are not available.

By building on existing frameworks such as the Time-Delta Method, this research leverages historical commit data and statistical models to estimate unobserved work periods. The Contribution Rate Imputation Method offers a more comprehensive view of developer productivity by focusing on both the size and complexity of code changes, enabling organizations to make more informed decisions regarding project timelines and resource management.

The objective of this paper is to present the theoretical foundation of CRIM, outline its methodological approach, and demonstrate its application through empirical analysis. This research aims to provide practical insights into how CRIM can be utilized to improve software development workflows and better understand the dynamics of developer effort in modern projects.

## 2 Background and Related Work

### 2.1 Existing Contribution Metrics

Measuring developer contributions to software repositories is a multifaceted challenge that has garnered significant attention in the field of software engineering. Various methods and metrics have been proposed to quantify the contributions of developers, each with its strengths and weaknesses. This synthesis will explore existing methods and metrics, drawing on a range of academic sources to provide a comprehensive overview.

One of the foundational approaches to measuring developer contributions is the classification of developers based on their activity levels within a repository. Costa et al. [1] categorize developers into three groups: core, active, and peripheral. This classification allows for a nuanced understanding of how different developers contribute to a project. Core developers are typically those who make the most significant contributions, while peripheral developers may contribute less frequently but can still play vital roles in specific contexts. The metrics used in this classification include code contributions, the frequency of buggy commits, and the resolution of priority bugs, which provide a quantitative basis for evaluating developer impact.

In addition to classification, repository mining-based metrics have been proposed to assess developer contributions. Lima et al. [2] emphasize the importance of mining version control systems to extract meaningful metrics that reflect developer activity. Metrics derived from repository mining include the number of commits, lines of code added or removed, and the complexity of changes made. These metrics can provide insights into not only the quantity of contributions but also their quality and impact on the overall project.

Another significant aspect of measuring contributions involves the analysis of pull requests (PRs) in collaborative environments, particularly in open-source projects. Şeker et al. [3] highlight the role of PRs as a key feature for developers to inform others about changes made to a repository. Metrics related to PRs, such as the number of PRs submitted, the time taken to merge them, and the feedback received, can serve as indicators of a developer's engagement and influence within a project.

Quality metrics also play a crucial role in assessing developer contributions. Bassi et al. [4] discuss the importance of integrating quality metrics into the evaluation of developer contributions. These metrics can include code quality assessments, such as cyclomatic complexity, code smells, and test coverage. By combining contribution metrics with quality metrics, a more holistic view of a developer's impact on software quality can be achieved.

The concept of ownership versus contribution has been explored to understand the alignment between a developer's ownership of code and their actual contributions. Zabardast et al. [5] propose a model that uses the Goal-Question-Metric (GQM) approach to investigate this alignment. This model allows for the identification of discrepancies between ownership and actual contributions, providing insights into how developer roles may evolve over time.

Social network analysis (SNA) has also emerged as a valuable tool for measuring developer contributions. Meneely et al. [6] advocate for the use of SNA to understand collaboration patterns among developers. By analyzing the interactions between developers, such as code reviews and issue resolutions, SNA can reveal the social dynamics that influence contributions. Metrics derived from SNA can include centrality measures, which indicate a developer's influence within the network, and clustering coefficients, which reflect the degree of collaboration among developers.

The use of entropy-based metrics has been proposed to quantify the diversity of contributions in open-source projects. Kazuki et al. [7] introduce an entropy-based metric that assesses the balance of developer contributions to a source file. This metric can indicate potential fault-proneness, as files with uneven contributions may be more susceptible to bugs. By analyzing the distribution of contributions, developers and project managers can identify areas that may require more balanced input from the team.

In educational contexts, measuring contributions can also be vital for assessing student performance in software development projects. Hamer et al. [8] discuss the application of Git metrics to evaluate students' contributions in team projects. Metrics such as commit frequency, lines of code, and participation in code reviews can provide educators with insights into individual and team performance, facilitating better feedback and support for students.

The integration of issue tracking annotations into developer activity metrics has been proposed to enhance the accuracy of contribution assessments. Meneely et al. [6] argue that incorporating issue tracking data can provide a clearer picture of a developer's contributions, as it links code changes to specific tasks or bugs. This approach allows for a more contextualized understanding of contributions, as it considers the relevance and impact of the work done.

The development of composite metrics that combine various dimensions of contribution has also been explored. Diamantopoulos et al. [9] emphasize the need for metrics that not only quantify contributions but also assess their quality and relevance to the project's goals. By integrating multiple metrics, such as code quality, contribution frequency, and issue resolution rates, a more comprehensive evaluation of developer contributions can be achieved.

In conclusion, measuring developer contributions to software repositories is a complex task that requires a multifaceted approach. The existing methods and metrics range from simple quantitative measures, such as commit counts and lines of code, to more sophisticated analyses involving quality metrics, social network analysis, and entropy-based assessments. By employing a combination of these methods, software teams can gain valuable insights into individual and collective contributions, ultimately leading to improved project outcomes and enhanced collaboration.

## 2.2  Time-Delta Method

The Contribution Rate Imputation Method (CRIM) builds on a foundational technique known as the Time-Delta Method, which provides a systematic framework for estimating software development contribution rates. The Time-Delta Method focuses on identifying time intervals between developer commits and associating these with contribution metrics such as code complexity and textual changes, thus allowing the derivation of a "contribution rate." Understanding this methodology is crucial to appreciating how CRIM estimates unobserved work periods.

### 2.2.1  Time-Delta Method Background

The Time-Delta Method, as developed by Bishop (2024) [10], quantifies a developer's effort by calculating the Commit Time Delta (CTD)—the elapsed time between consecutive commits made by the same author. By tracking the time gap (or delta) between an antecedent commit and a subsequent one, the method offers a reasonable proxy for estimating how long a developer might have worked on a given task.

One of the core elements of the Time-Delta Method is the use of software complexity metrics, such as Cyclomatic Complexity or Levenshtein distance. These metrics provide a quantifiable measure of the size and complexity of code changes made in each commit. The greater the complexity of a change, the more likely it represents a significant amount of developer effort. By combining these metrics with the CTD, the Time-Delta Method generates a Contribution Rate (CR), which serves as a per-hour measure of development work.

However, the Time-Delta Method has its limitations, especially when dealing with commits that span long periods of time or where continuous work may not have been observed. This is where the Contribution Rate Imputation Method (CRIM) comes into play, as it addresses the challenge of estimating developer effort during periods where direct observation or precise measurement is not possible.

### 2.2.2  The Role of the Time-Delta Method in CRIM

The CRIM extends the Time-Delta Method by incorporating statistical techniques to handle "missing" or unobserved work periods. By analyzing patterns in the CTD values and the corresponding contribution rates across a repository's history, CRIM can "impute" contribution rates for commits with extended or unusually long time gaps. This imputation is based on observed patterns in the data and employs models, such as Mean Bound Contribution Rate (MBCR), to estimate the developer's likely productivity during unmeasured periods.

For instance, commits with long CTD values (e.g., a week between two commits) may not necessarily represent a full week of continuous development effort. Using CRIM, contribution rates from shorter, more reliably observed periods are used to estimate how much of that week was actively spent working on the software. This helps to avoid overestimating developer productivity during prolonged CTD intervals.

The Time-Delta Method provides the necessary foundation for calculating contribution rates by linking time intervals between commits with software complexity metrics. CRIM builds on this framework by offering a way to estimate effort during unobserved or prolonged periods, thereby providing a more complete picture of developer productivity. This combination allows CRIM to improve upon traditional productivity metrics and offer better insights into how software projects can be managed and optimized.

## 2.3 Prior Work on Developer Effort Estimation

Estimating the time that a software developer spends on a commit from a software repository is a complex task that involves various methodologies and approaches. These methods can be broadly categorized into quantitative and qualitative analyses, each leveraging different aspects of commit data and developer behavior.

One prevalent quantitative method involves analyzing the timestamps associated with commits. Each commit in a version control system like Git is timestamped, which allows researchers to measure the time intervals between commits made by the same developer. This temporal analysis can reveal patterns of activity, such as peak working hours or periods of intense development activity. For instance, Zhang et al. highlight that the frequency of commits can reflect the active engagement of developers in their projects, suggesting that a higher frequency correlates with increased working time during specific periods [11]. Additionally, Eyolfson et al. explored correlations between commit characteristics and the likelihood of introducing bugs, indicating that time-based characteristics can provide insights into developer behavior and the quality of their commits [12].

Another quantitative approach is the use of commit message analysis. Commit messages often contain information about the nature and scope of changes made, which can be used to infer the time spent on specific tasks. Liu et al. emphasize the importance of high-quality commit messages for understanding the intent behind changes, suggesting that detailed messages may correlate with more substantial time investments in those changes [13]. Furthermore, the analysis of commit sizes—measured in lines of code added or removed—can also serve as a proxy for estimating the time spent on a commit. Larger commits may indicate more extensive changes that likely required more time to implement, as discussed by Eyolfson et al. in their investigations into commit bugginess and time-based characteristics [12].

Qualitative methods also play a crucial role in estimating developer time on commits. For instance, the analysis of developer collaboration patterns can provide insights into how time is allocated across different tasks. Gote et al. discuss how co-editing networks can be constructed from commit histories, revealing collaborative behaviors that may influence how developers allocate their time across various projects [14]. This collaborative aspect is further supported by findings from Kalliamvakou et al., who note that the structure of development teams and their interactions can significantly impact commit activities and, consequently, time estimation [15].

The use of machine learning techniques to analyze commit data has gained traction in recent years. By training models on historical commit data, researchers can predict the time spent on future commits based on patterns observed in previous behaviors. For example, Cho et al. propose extending developer experience metrics to improve just-in-time defect prediction, which inherently requires understanding the time dynamics associated with commits [16]. This predictive modeling can be particularly useful in large projects where manual tracking of developer time is impractical.

In addition to these methods, the integration of developer sentiment analysis into commit data can provide a more nuanced understanding of time allocation. Kuutila et al. suggest that emotional states reported by developers can correlate with their productivity levels, indicating that periods of high emotional engagement may align with increased time spent on commits [17]. This qualitative insight can complement quantitative measures, offering a holistic view of developer time management.

The analysis of commit dependencies can also inform time estimation. Dhaliwal et al. discuss how understanding the relationships between different commits can help identify which changes are related and how much time might have been spent on interconnected tasks [18]. This dependency analysis can be crucial in complex projects where changes are interdependent, thus complicating time estimation efforts.

The impact of external factors, such as the adoption of Continuous Integration (CI) practices, also influences commit behavior and time estimation. Baltes et al. found that CI can affect the frequency and nature of commits, suggesting that developers may adjust their commit strategies based on the CI environment, which in turn can impact time allocation [19]. This highlights the need for context-aware approaches when estimating time spent on commits.

The evolution of software projects over time can also affect how time is perceived and recorded. Benomar et al. discuss the phases of software evolution and how different development activities characterize these phases, which can provide insights into how time is distributed across various project stages [20]. Understanding these phases can help project managers better estimate the time developers might spend on future commits based on historical trends.

Estimating the time that a software developer spends on a commit involves a multifaceted approach that combines quantitative analyses of commit data, qualitative insights into developer behavior, and the integration of machine learning techniques. By leveraging various methodologies, including timestamp analysis, commit message evaluation, collaborative behavior analysis, and sentiment analysis, researchers can develop a more comprehensive understanding of developer time allocation. This holistic approach not only aids in time estimation but also enhances the overall management of software development processes.

## 3 Theory of Contribution Rate Imputation

The Contribution Rate Imputation Theory states:

> *"If the rate at which development commonly takes place is known, actual person hours for a software contribution can be estimated."*

### 3.1 Mathematical Foundation

The Contribution Rate Imputation Method (CRIM) is grounded in the principle that the effort expended on a software contribution can be estimated by analyzing historical contribution patterns. In CRIM, developer effort is modeled as a function of the size of a code contribution and the typical rate at which developers complete similar contributions. This estimation process allows us to impute the time spent on contributions, especially in cases where direct observation of time worked is not possible due to the asynchronous nature of modern development.

At its core, CRIM operates on two key metrics: the size of the contribution ($\Delta L$) and the model contribution rate ($\rho$). These metrics allow us to approximate the time ($\Delta t$) a developer spent on a particular task. The relationship between these variables is expressed mathematically as:

$$\Delta t = \frac{\Delta L}{\rho}$$

**Variables:**

- $\Delta t$: The estimated time spent on the contribution (in hours).
- $\Delta L$: The size of the contribution, which can be measured using metrics such as lines of code, cyclomatic complexity, or Levenshtein distance.
- $\rho$: The model contribution rate, which represents how much work a developer completes per unit of time, derived from historical data.

#### 3.1.1 Estimating Time ($\Delta t$)

The central output of the CRIM is the estimation of $\Delta t$, the time spent on a given contribution. By dividing the size of the contribution ($\Delta L$) by the model contribution rate ($\rho$), CRIM provides an estimate of the total time that a developer likely spent on the task. This imputation is especially useful in scenarios where developers work asynchronously or across distributed teams, and direct tracking of time spent is not possible.

This estimated time is not a direct measurement but rather an inference based on past behavior and contribution patterns. It allows project managers and developers to gain insights into how long tasks typically take and can inform resource allocation and project timelines.

### 3.1.2   Contribution Size (ΔL)

The size of the contribution, ΔL, is a critical component of this equation. It quantifies the amount of work represented by a commit or a series of commits. This can be measured using various complexity-based metrics such as:

- **Cyclomatic Complexity:** Captures the logical complexity of a contribution by measuring the number of independent paths through the code.
- **Levenshtein Distance:** Measures the textual difference between two versions of the same code, reflecting how much the source code has changed.
- **Lines of Code (LOC):** A simpler but less nuanced measure, which counts the lines added, removed, or modified.

While CRIM is adaptable to various contribution size metrics, cyclomatic complexity offers a novel approach by specifically measuring changes in the logic of code. This makes it particularly effective for contributions involving control structures, such as conditionals or loops. However, cyclomatic complexity is not applicable to code that lacks these structures, such as markup languages like HTML, where logical flow cannot be quantified. In such cases, using Levenshtein word distance provides a more language-agnostic alternative, allowing contributions to be measured regardless of the type of code. Additionally, Levenshtein distance enables a more familiar "words per minute" expression of contribution rate, which is more universally understood than metrics like "complexity delta per minute."

### 3.1.3   Model Contribution Rate ($\rho$)

The model contribution rate ($\rho$) represents a central element of the Contribution Rate Imputation Method (CRIM) and the CRIM is incomplete and not practically applicable without an effective mechanism to accurately identify model contribution rated. This section covers some considerations relevant to model contribution rates. A more in-depth and empirical analysis of candidate mechanisms for identifying model contribution rates is left as an opportunity for future research.

The contribution rate, $\rho$, is an empirical value that represents how quickly a developer typically completes contributions of a given size. The model contribution rate ($\rho$) serves as a proxy for the typical productivity of a developer when direct observation of development time is not feasible.

*Empirical Nature of $\rho$*

The model contribution rate is inherently empirical, meaning it must be derived from historical data and patterns of developer behavior. To accurately define $\rho$, large datasets of past contributions are analyzed, and a baseline for how much work is typically completed per unit of time is established. However, variations across developers, tasks, and even project types make this process challenging. Contributions can range in complexity from minor bug fixes to the development of new features, each requiring different levels of effort despite possibly appearing similar in size.

*Variability of Contributions*

One challenge in determining a reliable model contribution rate is the variability in the nature of the work being measured. Contributions are not uniform; they differ in complexity, structure, and the intellectual effort required. For example, a contribution of 100 lines of code that addresses a security vulnerability may demand more effort than a similar-sized contribution that adds a simple new feature. This variability must be accounted for when applying $\rho$, which may need to be adjusted for different types of contributions.

*Data-Driven Approaches to Estimate ρ*

Several data-driven approaches are employed to identify appropriate values for $\rho$. These approaches generally rely on large sets of historical data, examining patterns in the relationship between contribution size (measured in lines of code, cyclomatic complexity, or Levenshtein distance) and the time spent on those contributions. Some methods that can be used to derive the model contribution rate include:

- **Mean Bound Contribution Rate (MBCR):** This approach takes an average of the contribution rates observed in past work, specifically focusing on contributions that have known, measurable development times. The MBCR provides a baseline value for $\rho$, which can be applied to similar future contributions.
- **Regression Models:** More sophisticated models, such as Gradient Boosted Trees, can be used to estimate contribution rates by considering a range of factors that influence productivity, including the complexity of the code, the developer's past performance, and the type of project.

*Adjusting ρ for Context*

It could be necessary to adjust the model contribution rate to reflect the context in which it is being used. This may involve calibrating the rate based on factors such as:

- **Developer Experience:** More experienced developers may have higher contribution rates, as they are likely to work more efficiently on tasks they are familiar with.
- **Project Phase:** Early development phases (e.g., initial feature development) might have different rates compared to later stages (e.g., maintenance and bug fixes).
- **Task Complexity:** Highly complex contributions involving architectural changes or deep refactoring should have a lower $\rho$, acknowledging the greater intellectual effort required.

*Model Contribution Rate Challenges*

When the amount of data available is limited, determining an accurate model contribution rate ($\rho$) becomes significantly more challenging. A limited dataset may not capture the full range of contribution types, complexity levels, or developer behaviors, making it difficult to establish reliable baselines for productivity. Without sufficient historical data, the model contribution rate risks being skewed by outliers, such as unusually small or large contributions that do not represent typical development patterns. Additionally, small datasets may lack the statistical power needed to draw meaningful conclusions, leading to higher variability and lower confidence in the imputed values of $\rho$. This scarcity of data can also prevent the identification of important context-specific factors, such as the effect of developer experience or project phase, which are critical for adjusting $\rho$ accurately. As a result, the imputed contribution rates may be less reliable and could lead to inaccurate time estimates, limiting the effectiveness of the Contribution Rate Imputation Method.

*Limitations of the Model Contribution Rate*

While the model contribution rate is useful for estimating effort, it is not without limitations. One significant limitation is that it is an approximation based on past behavior, and it does not account for unpredictable factors such as interruptions, meetings, or other non-development activities that may affect actual work time. Additionally, the lack of real-time tracking means that $\rho$ can only provide an average estimate, which might not always be accurate for individual contributions.

The model contribution rate is an essential part of the Contribution Rate Imputation Theory, providing a mechanism to estimate development time when direct measurement is unavailable. However, its empirical nature and the variability of contributions necessitate careful consideration when applying it. Continued research and refinement of data-driven methods will improve the accuracy of $\rho$ and the effectiveness of CRIM in estimating developer productivity.

# 4 Practical Implications

The Contribution Rate Imputation Method (CRIM) is a practical tool designed to estimate developer effort by analyzing measurable contributions in software repositories. In environments where direct time tracking is limited or non-existent, CRIM provides an alternative, data-driven method to estimate time spent on development tasks. This section explores how CRIM can be applied in real-world scenarios to enhance project management and improve understanding of developer productivity.

## 4.1 Estimating Developer Effort

At its core, CRIM enables project managers to estimate the amount of time developers have spent on a specific task, even when traditional time-tracking methods are not used. This is particularly valuable in agile or fast-paced development environments, where contributions are made asynchronously, and detailed time logs may not be available. CRIM uses the size of the code contribution (in terms of lines of code, cyclomatic complexity, or Levenshtein distance) and the model contribution rate to impute the time taken by a developer to complete a contribution.

By doing so, project managers can more accurately gauge the effort required for specific contributions and understand how much time was likely spent on various parts of a project. This information provides insights into individual performance, helping teams evaluate whether developers are contributing efficiently relative to their historical performance or to the performance of their peers.

## 4.2 Application in Software Tools

CRIM can be integrated into software systems that analyze source code repositories, such as version control systems (e.g., Git). These tools can automatically analyze contributions over time and apply CRIM to estimate the hours spent on each contribution. For example, a product designed to examine source code repositories can generate reports showing the estimated time each developer spent on their commits. These reports can help project managers assess individual performance, compare contributions within teams, and identify patterns across repository cohorts.

This system offers project managers the ability to track performance trends across a repository or organization. For instance, a report might reveal that a particular developer or team consistently spends more time on complex features, providing opportunities for intervention, training, or adjustments to the project timeline. Additionally, the ability to track these metrics in real-time provides a powerful mechanism to fine-tune the allocation of resources and improve team productivity.

## 4.3 Optimizing Resource Allocation

CRIM's ability to estimate development time also aids in optimizing resource allocation. By understanding the estimated time spent on past contributions, project managers can allocate tasks more efficiently in the future. If certain developers are consistently completing contributions at a higher or lower rate than the average for their team, managers can adjust workloads to match individual strengths and weaknesses. This prevents underutilization or overburdening of developers, leading to a more balanced workload across the team.

The CRIM allows managers to forecast future project timelines with more accuracy. By identifying the average rate of contributions and estimating the time required for future tasks based on historical data, project managers can set more realistic deadlines. This ensures that projects stay on track and reduces the likelihood of delays caused by misjudged workloads.

## 4.4 Monitoring Productivity and Performance Trends

Beyond individual contributions, CRIM can also be used to monitor broader productivity trends within a repository or organization. By applying CRIM to past contributions, teams can track how the overall contribution rate has changed over time. This data provides valuable insights into how efficiently a team is working and whether improvements are needed.

For example, if a team's estimated contribution rates have decreased over the past few months, this may indicate issues such as burnout, overcomplicated tasks, or inefficient workflows. On the other hand, an increase in contribution rates could signal that team members are becoming more productive, possibly due to skill development, better tools, or more focused tasks.

## 4.5 Enhancing Decision-Making in Project Management

Ultimately, the practical application of CRIM enhances decision-making for project managers by offering a clearer picture of how development resources are being utilized. The insights gained through CRIM can help project managers decide where to focus their attention, whether it be reassigning tasks, adjusting timelines, or providing additional support to developers who may be struggling. The ability to estimate time spent on contributions with a high degree of accuracy allows for more data-driven decisions, making the development process more transparent and manageable.

In conclusion, CRIM is a powerful method for estimating developer effort, offering actionable insights for managing team productivity, optimizing resource allocation, and tracking long-term trends in software development environments. Integrating CRIM into systems that analyze source code repositories provides an effective way to monitor developer contributions and adjust project strategies to meet performance goals more effectively.

# 5 Future Work

The Contribution Rate Imputation Method (CRIM) has laid a theoretical and practical foundation for estimating developer effort in software development projects. However, its current implementation leaves room for enhancement, especially as technology evolves. Future research and development can address several areas to improve the accuracy, scalability, and application of CRIM in real-world environments. The following sections outline some of the key areas where future work can be focused.

## 5.1 Improving Model Contribution Rate Identification Using Artificial Intelligence

A critical challenge in the practical application of CRIM is the accurate identification of the model contribution rate ($\rho$). While historical data provides a basis for estimating $\rho$ leveraging artificial intelligence (AI) offers significant potential to enhance this process. AI and machine learning models can analyze large datasets from various development environments to uncover patterns that are difficult for traditional statistical methods to identify.

By training AI models on data such as commit history, contribution size, complexity, and developer productivity, it may be possible to create more accurate and context-aware models for $\rho$. These AI-driven models could automatically adjust for factors such as developer expertise, project type, and task complexity, refining the estimation process. Furthermore, AI can help detect anomalies in contribution patterns, ensuring that outliers do not disproportionately skew the model contribution rate. The use of AI also opens the possibility for real-time updates to $\rho$, allowing CRIM to evolve dynamically as new contributions are made. This would significantly improve the adaptability and accuracy of CRIM in rapidly changing development environments.

## 5.2 Enhancements to the CRIM Model

Beyond AI-driven methods, there are other enhancements that could be made to the CRIM model itself. For example, future research could explore new metrics for measuring contribution size beyond traditional lines of code, cyclomatic complexity, and Levenshtein distance. These new metrics could capture additional dimensions of developer effort, such as the intellectual complexity of tasks or the degree of collaboration required.

Moreover, the CRIM model could be expanded to account for non-code contributions that are often critical to software development but are harder to quantify, such as code reviews, bug triaging, and documentation. These activities, though not directly reflected in commit history, have a significant impact on project progress and developer workload. Integrating them into CRIM would provide a more holistic view of developer effort and contribution.

## 5.3 Broader Applications of CRIM

While CRIM is currently applied to software development environments, there are opportunities to expand its application into other areas where contributions need to be estimated based on limited data. For example, CRIM could be adapted to estimate effort in research and development projects, creative industries, or any environment where collaborative work is tracked through digital contributions.

In addition, CRIM's principles could be extended to interdisciplinary teams where contributions come from diverse fields (e.g., designers, product managers, marketers) working together on a shared project. In such environments, defining a generalized contribution metric that accounts for different types of work could enhance the applicability of CRIM.

## 5.4 Integration with Project Management Tools

For CRIM to achieve its full potential in real-world applications, it must be seamlessly integrated with popular project management tools and version control systems. Tools like Jira, GitLab, and GitHub already track developer activity and project progress. By embedding CRIM into these platforms, it would be possible to automate the analysis of contribution rates and generate real-time insights about developer productivity and effort.

Integration with project management tools also opens up opportunities for more intelligent scheduling and workload balancing. For example, CRIM could help predict how long it will take to complete certain tasks based on historical data, allowing project managers to plan more accurately. It could also be used to flag when developers are under- or over-performing, enabling more proactive interventions to maintain balanced workloads.

# 6 Conclusions

The Contribution Rate Imputation Method (CRIM) presents a significant advancement in software engineering, specifically in the domain of developer effort estimation. This method goes beyond traditional productivity metrics by providing a structured, data-driven approach to analyzing developer contributions through commit history, complexity measures, and statistical imputation. The following sections outline the scientific contributions that CRIM offers.

## 6.1 A Novel Framework for Effort Estimation

CRIM introduces a comprehensive framework for estimating developer effort in the absence of direct observation. By building on the Time-Delta Method, CRIM uses a combination of historical commit data and complexity metrics to estimate contribution time more accurately. This imputation approach bridges the gap between theoretical models and real-world developer behavior by accounting for unobserved work periods.

The key contribution here is the method's ability to model effort as a function of the size and complexity of contributions, transforming raw commit data into actionable insights. The introduction of CRIM thus provides a more precise tool for researchers studying software productivity, especially in distributed and asynchronous environments where developer activity is hard to track.

## 6.2 Advancements in Complexity and Contribution Size Metrics

CRIM builds upon existing complexity metrics like cyclomatic complexity and Levenshtein distance, incorporating them into a unified model that quantifies the effort required for specific contributions. By employing these measures, CRIM provides a nuanced understanding of developer contributions that transcends traditional metrics such as lines of code.

The scientific advancement here lies in CRIM's ability to quantify logical and textual changes in code, which allows it to reflect not just the quantity but the quality and difficulty of the contributions. This more granular assessment of developer effort provides a richer dataset for future research in software engineering, enabling deeper analysis of the relationship between code complexity and developer productivity.

### 6.3 Statistical Imputation of Developer Effort

One of the core innovations of CRIM is its use of statistical imputation to estimate unobserved work periods. By analyzing historical contribution patterns, CRIM can impute the likely amount of effort spent on contributions that lack direct time-tracking data. The inclusion of the Mean Bound Contribution Rate (MBCR) and other statistical models introduces a level of rigor to developer productivity analysis that has not been commonly applied in prior methods.

This use of imputation techniques allows CRIM to handle scenarios where developer activity spans long intervals between commits, ensuring that effort estimations are not skewed by outliers or periods of inactivity. This statistical innovation enhances the accuracy of software effort estimation, making it a valuable contribution to empirical software engineering research.

### 6.4 Expansion of Developer Effort Research

CRIM expands the scope of developer effort research by providing a conceptual model that can be applied across a variety of projects and development environments. Its flexible architecture allows for the integration of additional data, such as developer experience, project phase, or non-code contributions, which opens new avenues for future research in both academic and industrial contexts.

The method's adaptability to different types of contributions—whether code-related or otherwise—suggests that CRIM could be employed beyond traditional software engineering, perhaps in other collaborative environments where effort estimation is crucial. This interdisciplinary potential positions CRIM as a foundational model for future work on productivity and effort estimation in complex, collaborative domains.

### 6.5 Data-Driven Refinement of Contribution Rates

CRIM's model contribution rate ($\rho$) is a critical element, and its empirical nature adds a new dimension to research on developer productivity. While traditional models rely on fixed rates or averages, CRIM enables researchers to derive $\rho$ dynamically from large datasets of historical contributions. This data-driven refinement allows for more accurate predictions of developer effort, accounting for variations in task complexity, developer experience, and project phase.

The introduction of empirical, data-driven contribution rates opens the door to further refinement through artificial intelligence and machine learning techniques. As these technologies evolve, CRIM can be enhanced to provide even more precise estimates of developer effort, setting the stage for cutting-edge research in software project management and productivity analysis.

### 6.6 Contributions to Quantitative and Qualitative Software Engineering Research

CRIM bridges the gap between quantitative and qualitative research in software engineering. On the quantitative side, its reliance on numerical complexity metrics and commit data makes it a powerful tool for large-scale, data-driven studies of developer productivity. On the qualitative side, its focus on the intellectual and collaborative aspects of coding provides insights into how different types of contributions are made and the time they require.

This balanced approach allows CRIM to contribute to multiple streams of software engineering research, particularly in the study of developer behavior, collaboration patterns, and the relationship between task complexity and time investment. By integrating both quantitative and qualitative elements, CRIM provides a holistic framework for understanding developer effort in modern software projects.

## Acknowledgements

The authors would like to express their deepest gratitude to the Systems Engineering Department at Colorado State University, Fort Collins, for their invaluable support in making this research possible.

*During the preparation of this work, artificial intelligence was used in the research and writing processes. After using these services, the content was reviewed and edited as needed and full responsibility is taken for the content of this work.*

*This research predates Bishop's employment with Amazon, Inc.*